\PassOptionsToPackage{table,xcdraw}{xcolor}

\documentclass[sigconf]{acmart}

\AtBeginDocument{%
  \providecommand\BibTeX{{%
    \normalfont B\kern-0.5em{\scshape i\kern-0.25em b}\kern-0.8em\TeX}}}

\usepackage{amsthm}
\usepackage{bm}
\usepackage[table,xcdraw]{xcolor}
\usepackage{multicol}
\usepackage{colortbl}
\usepackage{amsmath}
\usepackage{multirow}
\usepackage[normalem]{ulem}
\useunder{\uline}{\ul}{}
\usepackage{hyperref}
\usepackage{graphicx}
\usepackage{subfigure}
\usepackage{enumitem}
\usepackage{threeparttable}  
\usepackage{booktabs}
\usepackage{bbding}
\usepackage{color}
\usepackage{xcolor}

\usepackage{algorithm}
\usepackage{listings}
\usepackage{etoolbox}
\makeatletter
\AfterEndEnvironment{algorithm}{\let\@algcomment\relax}
\AtEndEnvironment{algorithm}{\kern2pt\hrule\relax\vskip3pt\@algcomment}
\let\@algcomment\relax
\newcommand\algcomment[1]{\def\@algcomment{\footnotesize#1}}
\renewcommand\fs@ruled{\def\@fs@cfont{\bfseries}\let\@fs@capt\floatc@ruled
  \def\@fs@pre{\hrule height.8pt depth0pt \kern2pt}%
  \def\@fs@post{}%
  \def\@fs@mid{\kern2pt\hrule\kern2pt}%
  \let\@fs@iftopcapt\iftrue}
\makeatother

\setcopyright{none}

\acmSubmissionID{1573}

\copyrightyear{2022}
\acmYear{2022}
\setcopyright{acmlicensed}
\acmConference[WWW '22]{Proceedings of the ACM Web Conference 2022}{April 25--29, 2022}{Virtual Event, Lyon, France} \acmBooktitle{Proceedings of the ACM Web Conference 2022 (WWW '22), April 25--29, 2022, Virtual Event, Lyon, France}
\acmPrice{15.00}
\acmDOI{10.1145/3485447.3512022} 
\acmISBN{978-1-4503-9096-5/22/04}

\begin{document}

\fancyhead{}

\title{Hybrid Contrastive Quantization for \\Efficient Cross-View Video Retrieval}

\author{Jinpeng Wang$^{1,3\dagger}$, Bin Chen$^{2\ddag}$, Dongliang Liao$^{4\ddag}$, Ziyun Zeng$^{1,3}$, Gongfu Li$^{4}$, Shu-Tao Xia$^{1,3}$, Jin Xu$^{4}$
}

\affiliation{
    \country{$^1$ Tsinghua Shenzhen International Graduate School, Tsinghua University} \qquad
    \country{$^2$ Harbin Institute of Technology, Shenzhen,}\\
    \country{$^3$ Research Center of Artificial Intelligence, Peng Cheng Laboratory}\qquad
    \country{$^4$ Data Quality Team, Wechat Group, Tencent Inc.}
}
\thanks{$\dagger$ Supported by 2021 Tencent Rhino-Bird Research Elite Training Program.}
\thanks{$\ddag$ Corresponding authors.}

\email{wjp20@mails.tsinghua.edu.cn, chenbin2021@hit.edu.cn, brightliao@tencent.com, zengzy21@mails.tsinghua.edu.cn,}
\email{gongfuli@tencent.com, xiast@sz.tsinghua.edu.cn, cnjinxu@gmail.com}


\definecolor{ForestGreen}{RGB}{34,139,34}

\newcommand{\ie}{\emph{i.e.},~}
\newcommand{\eg}{\emph{e.g.},~}
\newcommand{\wrt}{\emph{w.r.t.}~}
\newcommand{\aka}{\emph{aka}~}
\renewcommand{\paragraph}[1]{\medskip\noindent\textbf{#1.~}}
\newcommand{\todo}{\color{red}{\textbf{TODO.}}~}

\newcommand{\xiaok}[1]{\left(#1\right)}
\newcommand{\zhongk}[1]{\left[#1\right]}
\newcommand{\dak}[1]{\left\{#1\right\}}
\newcommand{\jiaok}[1]{\left<#1\right>}
\newcommand{\shuk}[1]{\left\lVert#1\right\rVert}
\newcommand{\shuks}[1]{\left\lVert#1\right\rVert^2}
\newcommand{\shangk}[1]{\left\lceil #1 \right\rceil}
\newcommand{\xiak}[1]{\left\lfloor #1 \right\rfloor}

\newcommand{\argmax}[1]{{\mathop{\arg\mathrm{max}}_{#1}\,}}
\newcommand{\argmin}[1]{{\mathop{\arg\mathrm{min}}_{#1}\,}}
\newcommand{\pfrac}[2]{\frac{\partial #1}{\partial #2}}
\newcommand{\prob}[2]{p\xiaok{#1 \mid #2}}

\newcommand{\T}{\top}
\newcommand{\dif}{\mathop{}\!\mathrm{d}}
\newcommand{\biset}[1]{\{0,1\}^{#1}}
\newcommand{\ReLU}{\mathrm{ReLU}}
\newcommand{\relu}{\mathrm{ReLU}}
\newcommand{\sign}{\mathrm{sign}}
\newcommand\softmax{\mathrm{softmax}}
\newcommand\KL{D_{\mathrm{KL}}}
\newcommand\Var{\mathrm{Var}}
\newcommand\Cov{\mathrm{Cov}}
\newcommand{\Tr}{\mathrm{Tr}}
\newcommand{\tr}{\mathrm{tr}}
\newcommand{\dist}{\mathrm{dist}}
\newcommand{\concat}{\mathrm{concat}}
\newcommand{\mean}{\mathrm{mean}}
\newcommand{\diag}{\mathrm{diag}}
\newcommand{\cov}{\mathrm{cov}}

\newcommand{\range}[1]{{0,1,\cdots,#1}} 
\newcommand{\Range}[1]{{1,2,\cdots,#1}} 
\newcommand{\opseq}[3]{{#1_1 #3 #1_2 #3 \cdots #3 #1_{#2}}}
\newcommand{\seq}[2]{\opseq{#1}{#2}{,}}
\newcommand{\xseq}[2]{\opseq{#1}{#2}{\times}}

\newcommand{\bma}{\bm{a}}
\newcommand{\bmb}{\bm{b}}
\newcommand{\bmc}{\bm{c}}
\newcommand{\bmd}{\bm{d}}
\newcommand{\bme}{\bm{e}}
\newcommand{\bmf}{\bm{f}}
\newcommand{\bmg}{\bm{g}}
\newcommand{\bmh}{\bm{h}}
\newcommand{\bmi}{\bm{i}}
\newcommand{\bmj}{\bm{j}}
\newcommand{\bmk}{\bm{k}}
\newcommand{\bml}{\bm{l}}
\newcommand{\bmm}{\bm{m}}
\newcommand{\bmn}{\bm{n}}
\newcommand{\bmo}{\bm{o}}
\newcommand{\bmp}{\bm{p}}
\newcommand{\bmq}{\bm{q}}
\newcommand{\bmr}{\bm{r}}
\newcommand{\bms}{\bm{s}}
\newcommand{\bmt}{\bm{t}}
\newcommand{\bmu}{\bm{u}}
\newcommand{\bmv}{\bm{v}}
\newcommand{\bmw}{\bm{w}}
\newcommand{\bmx}{\bm{x}}
\newcommand{\bmy}{\bm{y}}
\newcommand{\bmz}{\bm{z}}
\newcommand{\bmzero}{\bm{0}}
\newcommand{\bmone}{\bm{1}}
\newcommand{\bmalpha}{\bm{\alpha}}
\newcommand{\bmbeta}{\bm{\beta}}
\newcommand{\bmgamma}{\bm{\gamma}}
\newcommand{\bmdelta}{\bm{\delta}}
\newcommand{\bmepsilon}{\bm{\epsilon}}
\newcommand{\bmtheta}{\bm{\theta}}
\newcommand{\bmiota}{\bm{\iota}}
\newcommand{\bmkappa}{\bm{\kappa}}
\newcommand{\bmlambda}{\bm{\lambda}}
\newcommand{\bmmu}{\bm{\mu}}
\newcommand{\bmnu}{\bm{\nu}}
\newcommand{\bmxi}{\bm{\xi}}
\newcommand{\bmpi}{\bm{\pi}}
\newcommand{\bmrho}{\bm{\rho}}
\newcommand{\bmsigma}{\bm{\sigma}}
\newcommand{\bmtau}{\bm{\tau}}
\newcommand{\bmupsilon}{\bm{\upsilon}}
\newcommand{\bmphi}{\bm{\phi}}
\newcommand{\bmchi}{\bm{\chi}}
\newcommand{\bmpsi}{\bm{\psi}}
\newcommand{\bmomega}{\bm{\omega}}
\newcommand{\bmA}{\bm{A}}
\newcommand{\bmB}{\bm{B}}
\newcommand{\bmC}{\bm{C}}
\newcommand{\bmD}{\bm{D}}
\newcommand{\bmE}{\bm{E}}
\newcommand{\bmF}{\bm{F}}
\newcommand{\bmG}{\bm{G}}
\newcommand{\bmH}{\bm{H}}
\newcommand{\bmI}{\bm{I}}
\newcommand{\bmJ}{\bm{J}}
\newcommand{\bmK}{\bm{K}}
\newcommand{\bmL}{\bm{L}}
\newcommand{\bmM}{\bm{M}}
\newcommand{\bmN}{\bm{N}}
\newcommand{\bmO}{\bm{O}}
\newcommand{\bmP}{\bm{P}}
\newcommand{\bmQ}{\bm{Q}}
\newcommand{\bmR}{\bm{R}}
\newcommand{\bmS}{\bm{S}}
\newcommand{\bmT}{\bm{T}}
\newcommand{\bmU}{\bm{U}}
\newcommand{\bmV}{\bm{V}}
\newcommand{\bmW}{\bm{W}}
\newcommand{\bmX}{\bm{X}}
\newcommand{\bmY}{\bm{Y}}
\newcommand{\bmZ}{\bm{Z}}
\newcommand{\bmGamma}{\bm{\Gamma}}
\newcommand{\bmDelta}{\bm{\Delta}}
\newcommand{\bmTheta}{\bm{\Theta}}
\newcommand{\bmLambda}{\bm{\Lambda}}
\newcommand{\bmXi}{\bm{\Xi}}
\newcommand{\bmPi}{\bm{\Pi}}
\newcommand{\bmSigma}{\bm{\Sigma}}
\newcommand{\bmUpsilon}{\bm{\Upsilon}}
\newcommand{\bmPhi}{\bm{\Phi}}
\newcommand{\bmPsi}{\bm{\Psi}}
\newcommand{\bmOmega}{\bm{\Omega}}

\newcommand{\calA}{\mathcal{A}}
\newcommand{\calB}{\mathcal{B}}
\newcommand{\calC}{\mathcal{C}}
\newcommand{\calD}{\mathcal{D}}
\newcommand{\calE}{\mathcal{E}}
\newcommand{\calF}{\mathcal{F}}
\newcommand{\calG}{\mathcal{G}}
\newcommand{\calH}{\mathcal{H}}
\newcommand{\calI}{\mathcal{I}}
\newcommand{\calJ}{\mathcal{J}}
\newcommand{\calK}{\mathcal{K}}
\newcommand{\calL}{\mathcal{L}}
\newcommand{\calM}{\mathcal{M}}
\newcommand{\calN}{\mathcal{N}}
\newcommand{\calO}{\mathcal{O}}
\newcommand{\calP}{\mathcal{P}}
\newcommand{\calQ}{\mathcal{Q}}
\newcommand{\calR}{\mathcal{R}}
\newcommand{\calS}{\mathcal{S}}
\newcommand{\calT}{\mathcal{T}}
\newcommand{\calU}{\mathcal{U}}
\newcommand{\calV}{\mathcal{V}}
\newcommand{\calW}{\mathcal{W}}
\newcommand{\calX}{\mathcal{X}}
\newcommand{\calY}{\mathcal{Y}}
\newcommand{\calZ}{\mathcal{Z}}

\newcommand{\bbC}{\mathbb{C}}
\newcommand{\bbE}{\mathbb{E}}
\newcommand{\bbN}{\mathbb{N}}
\newcommand{\bbQ}{\mathbb{Q}}
\newcommand{\bbR}{\mathbb{R}}
\newcommand{\bbZ}{\mathbb{Z}}

\newcommand{\tabincell}[2]{\begin{tabular}{@{}#1@{}}#2\end{tabular}}

\def \CBNS {CBNS}
\begin{abstract}
With the recent boom of video-based social platforms (\eg YouTube and TikTok), video retrieval using sentence queries has become an important demand and attracts increasing research attention. 
Despite the decent performance, existing text-video retrieval models in vision and language communities are impractical for large-scale Web search because they adopt brute-force search based on high-dimensional embeddings. 
To improve efficiency, Web search engines widely apply vector compression libraries (\eg FAISS~\cite{FAISS}) to post-process the learned embeddings.
Unfortunately, separate compression from feature encoding degrades the robustness of representations and incurs performance decay. 
To pursue a better balance between performance and efficiency, we propose the first quantized representation learning method for cross-view video retrieval, namely \textbf{H}ybrid \textbf{C}ontrastive \textbf{Q}uantization (\textbf{HCQ}). 
Specifically, HCQ learns both \emph{coarse-grained} and \emph{fine-grained} quantizations with transformers, which provide complementary understandings for texts and videos and preserve comprehensive semantic information. 
By performing Asymmetric-Quantized Contrastive Learning (AQ-CL) across views, HCQ aligns texts and videos at coarse-grained and multiple fine-grained levels. 
This hybrid-grained learning strategy serves as strong supervision on the cross-view video quantization model, where contrastive learning at different levels can be mutually promoted. 
Extensive experiments on three Web video benchmark datasets demonstrate that HCQ achieves competitive performance with state-of-the-art non-compressed retrieval methods while showing high efficiency in storage and computation.
\end{abstract}

%
%

%
%


%


\maketitle

\section{Introduction}
\label{sec: introduction}

\begin{figure}[t]
    \centering
    \includegraphics[width=0.87\columnwidth]{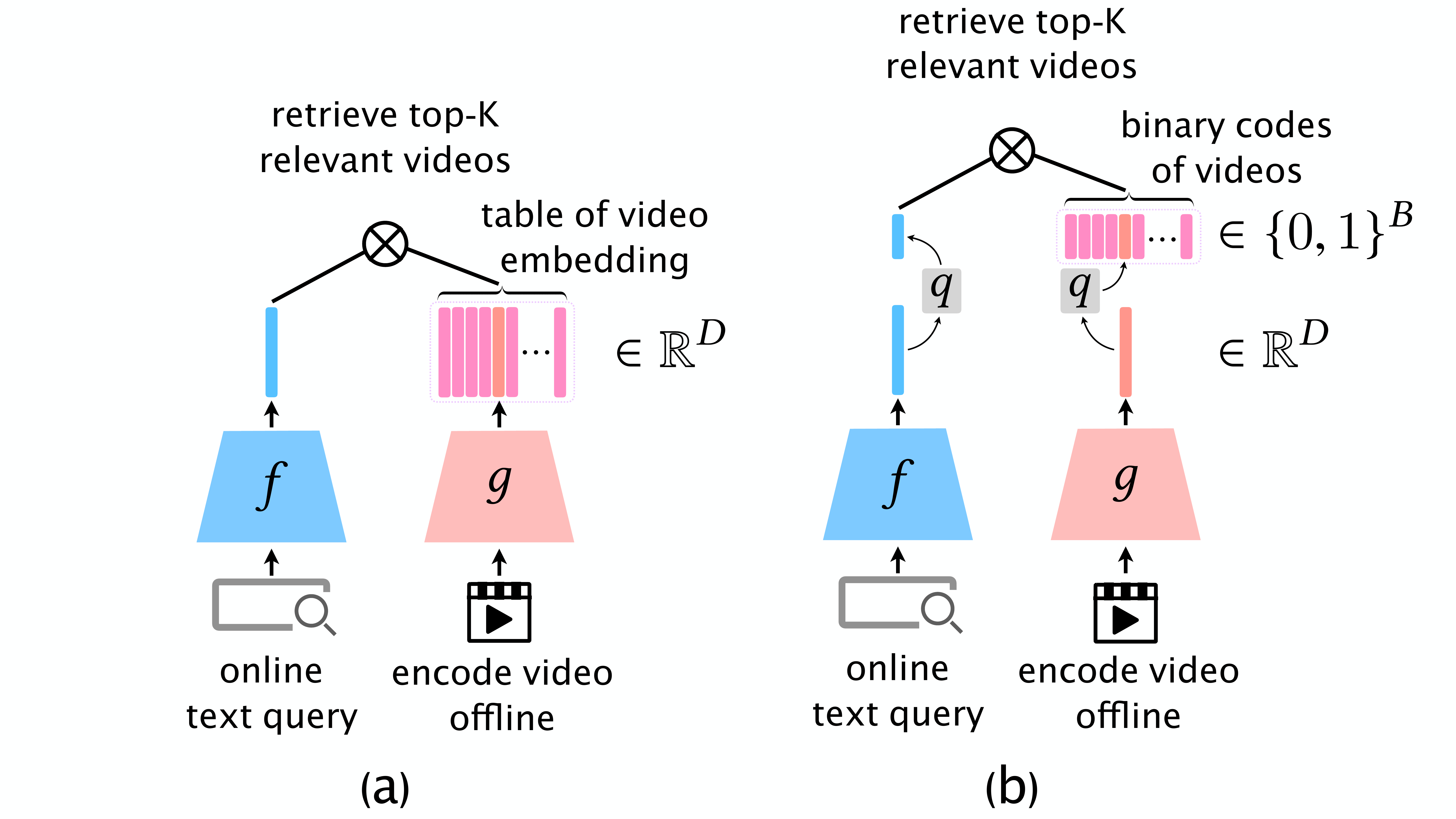}
    \caption{
    (a) The widely-adopted dual-encoder architecture in cross-view video retrieval. 
    (b) To facilitate search efficiency, practical Web search engines apply independent compression (denoted by $q$) to post-process embeddings, which can incur large performance decay. 
    In contrast, this paper delves into jointly learning dual-encoder and vector quantization, \ie jointly learning $f$, $g$ and $q$.
    }
    \label{fig:intro}
\end{figure}

\begin{figure}[t]
    \centering
    \includegraphics[width=\columnwidth]{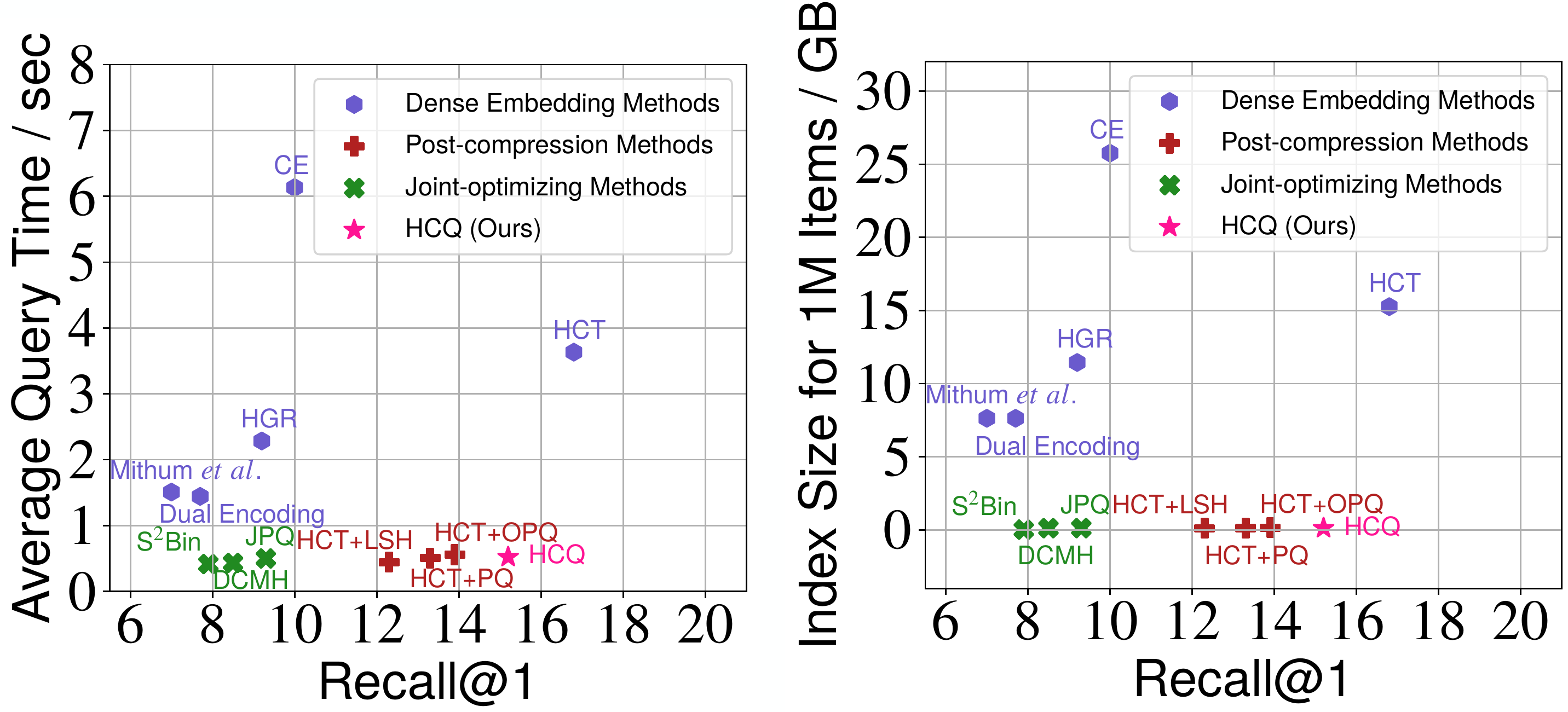}
    \caption{Model performance (recall@1) versus model efficiency (left: processing time per text-to-video query; right: storage overhead with a database of 1M videos) on MSRVTT dataset~\cite{MSRVTT} (``Full'' split). The proposed HCQ reaches a good tradeoff between the performance and the efficiency.}
    \label{fig:perf_eff}
\end{figure}

The temporal dynamics and abundant multi-modal content make video one of the most informative media to humans. 
As videos are gaining in popularity along with social media (\eg Facebook and Twitter) and video sharing platforms (\eg YouTube, Instagram and TikTok), cross-view (\eg sentence-based) video retrieval becomes an important demand for Web search. 
In vision and language research fields, many efforts have been devoted to text-video retrieval. 
Prior arts~\cite{CCA,JSFusion,HSE,CE,JPose,DualEncoding,HGR,MMT,COOT,T2VLAD,HANet} delve into dual-encoder architectures (as shown in  Fig.\ref{fig:intro}(a)) that produce separate text and video embeddings for dense retrieval. 
Despite the promising performance, these methods adopt brute-force search with high-dimensional embeddings, which is prohibitive in practical Web search due to the efficiency concern\footnote{As shown in Fig.\ref{fig:perf_eff}, though HCT (a dense-embedding-based variant of the proposed HCQ) yields the best performance, it requires $\sim$4 seconds to process a single query.}. 

Large-scale cross-view video retrieval is a challenging task, owing to the struggle between performance and efficiency. 
Hashing~\cite{L2H-1,L2H-2} is a common choice to facilitate the efficiency of dense retrieval, which generally contains two categories, \ie \emph{binary hashing} and \emph{quantization}. 
Binary hashing methods \cite{LSH,SH,ITQ,ADSH,DPSH,HashNet,DSH} transform data into the Hamming space such that distance can be measured with fast bitwise operations. 
Quantization methods \cite{PQ,OPQ,AQ,DTQ,DQN,PQN,WSDAQ} divide real data space into disjoint clusters and approximately represent the data points in each cluster as its centroid. 
Pre-computing inter-centroid distances in a lookup table can largely accelerate distance computation between quantized vectors. 
As hashing algorithms have been implemented by many Approximate Nearest Neighbor Search (ANNS) libraries (\eg FAISS~\cite{FAISS} and Annoy~\cite{Annoy}), practical Web search engines widely apply them to post-process embeddings (as shown in Fig.\ref{fig:intro}(b)). 
However, independent compression on dense embeddings can lead to large semantic deviation, which degrades the representation robustness and incurs performance decay. 

To mitigate the performance decay, one promising direction is to learn embedding and compression jointly. 
Such ideas have been explored in related fields like text embedding quantization~\cite{Poeem,JPQ,MoPQ,RepCONC} and cross-modal image hashing~\cite{DVSH,PRDH,DCMH,CDQ,HiCHNet,HSCH}, demonstrating improvements over post-compression. 
It is reasonable to extend some of them to our task, for example, by replacing the image encoder in cross-modal hashing models with a video encoder. 
Nevertheless, such immediate extensions are still far from enough in cross-view video retrieval. 
Note that text quantization and cross-modal image hashing take either the output of global pooling layer or the condensed token to generate binary codes. 
In this way, they can only preserve limited semantic information from the input and support coarse-grained cross-view alignment. 
Compared with texts and images, videos contain richer multi-modal information and temporal dynamics that are challenging to model. 
Thus, representing single vectors and simply focusing on coarse-grained alignment will increase the learning difficulty and result in insufficient representations. 
Unfortunately, there are few hashing models tailored for cross-view video retrieval nor suitable hashing methods considering fine-grained representation, leaving the problem of cross-view video hashing under-explored.

To overcome the defect, we present the first quantized representation learning method for efficient cross-view video retrieval, namely \textbf{H}ybrid \textbf{C}ontrastive \textbf{Q}uantization (\textbf{HCQ}). 
Without the reliance on hand-crafted hierarchical representations or complex graph reasoning~\cite{HGR,ViSENR,HANet}, it learns quantized representations through an end-to-end framework. 
Specifically, HCQ takes advantage of transformers~\cite{vaswani2017attention} to encode text features and multi-modal video features as bags of token embeddings, based on which coarse-grained and fine-grained semantic information is fully exploited. 
For the \emph{coarse-grained} representations, HCQ condenses each view into a global output. 
For the \emph{fine-grained} representations, HCQ further learns a common space with local semantic structure via a GhostVLAD~\cite{GhostVLAD} module. 
In this space, token embeddings of each view are softly assigned to a set of parameterized clusters, each of which stands for a latent topic or semantic concept. 
Learnable quantization is applied to the residual aggregations, producing fine-grained representations with complementary understanding. 
Moreover, we design a hybrid contrastive learning strategy that aligns text and video views at coarse-grained and multiple fine-grained levels. 
At each level, HCQ performs Asymmetric-Quantized Contrastive Learning (AQ-CL) across the two views. 
This strategy serves as strong supervision on cross-view video quantization model, where contrastive alignments at different levels are mutually promoted. 

To summarize, we make the following contributions.
\setlist{nolistsep}
\begin{itemize}[leftmargin=1.5em]
	\item We present the first quantization learning method for cross-view video retrieval, namely Hybrid Contrastive Quantization. 
	\item We proposed to learn both \emph{coarse-grained} and \emph{fine-grained} quantizations that preserve comprehensive semantic information. 
	\item We design a hybrid contrastive learning strategy to align texts and videos at coarse-grained and multiple fine-grained levels. It serves as strong supervision on the cross-view video quantization model and facilitates mutual promotion across levels. 
	\item Extensive experiments demonstrate that our method is competitive against state-of-the-art non-compressed methods while showing high efficiency in storage and computation.
\end{itemize}

\section{Related Work}
\label{sec:related_work}

\subsection{Cross-View Video Retrieval}
\label{subsec:vid_retr}
Cross-view (\eg sentence-based) video retrieval plays an essential role in the Web search.
Extensive efforts in computer vision and information retrieval have been made towards dual-encoder designs and cross-view matching schemes.
Simple methods~\cite{JSFusion,W2VV2,VSE2,mithun2018learning,HowTo100M,DualEncoding,TreeAug} modeled sequential information of videos and texts by bag-of-words representation, basic pooling, convolution modules, recurrent networks, or their combinations. 
To further refine the feature encoding, \citet{MoEE,CE,MMT} exploited multi-modal cues, \eg motion, appearance, audio, speech, to comprehensively represent videos. 
These methods also leveraged Mixture-of-Expert (MoE) strategies to match two views with multiple modalities. 
From another perspective of fine-grained or hierarchical representations, a series of methods exploited extra tools (\eg object detectors and sentence parsers) to pre-process inputs and handcrafted multi-level semantic information. 
For example, \citet{JPose} extracted the nouns and verbs in the sentences and projected them into a shared space with videos. 
Graph-based methods~\cite{HGR,ViSENR,HCGC,HANet} decomposed text or/and video inputs into different semantic roles (\eg entities, actions and events), based on which they performed delicate graph reasoning to refine the representations. 
Recently, dual-transformer-based methods~\cite{MMT,T2VLAD,HiT} have attracted increasing attention, not only for their brief and unified manners but also due to superior performance. 
Nevertheless, they only focused on coarse-grained representation and weakly aligned texts and videos by pairwise ranking objectives.

Note that the above methods produce high-dimensional embeddings and adopt brute-force computation for cross-view matching. 
Thus they are prohibitive in practical Web search due to the efficiency concern. 
On the other hand, learning efficient representations for cross-view video retrieval is still under-explored. 
To our best knowledge, there is only \emph{one} method termed S$^2$Bin~\cite{qi2021semantics}, which is a binary hashing method on this topic. 
It designed a complex system, including a spatial-temporal encoder-decoder for video representation, a convolutional text encoder, and an attribute network to predict categorical information for detected region proposals.

Unlike existing works, this paper contributes the first quantization learning framework, namely Hybrid Contrastive Quantization (HCQ), for efficient cross-view video retrieval.
HCQ exploits the robust dual-transformer architecture~\cite{MMT,T2VLAD,HiT} for encoding text features and multi-modal video features, which provides two advantages over S$^2$Bin~\cite{qi2021semantics}:
(\textbf{i}) HCQ can leverage multi-modal cues to learn video representations, preserving more comprehensive semantic information.
(\textbf{ii}) HCQ can effectively learn coarse-grained and fine-grained representations without requiring explicit supervision from the attribute detection module.
In HCQ, the contrastive learnings at coarse-grained and multiple fine-grained levels serve as solid supervision and improve the representations effectively.

\subsection{Learning to Hash}
\label{subsec:l2h}
Hashing~\cite{L2H-1,L2H-2} is a common choice to accelerate dense retrieval (\aka the \emph{first-stage retrieval} in information retrieval). 
Traditional hashing includes binary hashing~\cite{LSH,SH,SphH,ITQ,AGH} and quantization~\cite{PQ,OPQ,LOPQ,AQ,CQ}. 
Some of these methods, \eg LSH~\cite{LSH} and PQ~\cite{PQ}, have been widely implemented in Approximate Nearest Neighbor Search (ANNS) libraries, \eg FAISS~\cite{FAISS} and Annoy~\cite{Annoy}, and are adopted by search engines to post-process embeddings. 
Nevertheless, independent compression after representation learning can lead to significant semantic deviation, degrading the representation robustness and incurring performance decay. 
In contrast, jointly learning embedding and compression can improve the quality of final representations and mitigate the performance decay. 
Recently, S$^2$Bin~\cite{qi2021semantics} first took this idea for cross-view video retrieval, as discussed in the previous section (\S~\ref{subsec:vid_retr}). 
Besides, other hashing applications like text embedding quantization\cite{Poeem,JPQ,MoPQ,RepCONC} and cross-modal image hashing~\cite{DVSH,PRDH,DCMH,CDQ,HiCHNet,HSCH} have explored similar ideas, showing considerable improvements over post-compression. 

Our Hybrid Contrastive Quantization (HCQ) is the first method of cross-view video quantization learning. 
Besides coarse-grained quantization, we also propose to learn fine-grained quantization. 
We learn this hybrid-grained quantization by Asymmetric-Quantized Contrastive Learning (AQ-CL), which is tailored for cross-view video retrieval and has yet been studied in previous works. 
\begin{figure*}[t]
    \centering
    \includegraphics[width=0.86\textwidth]{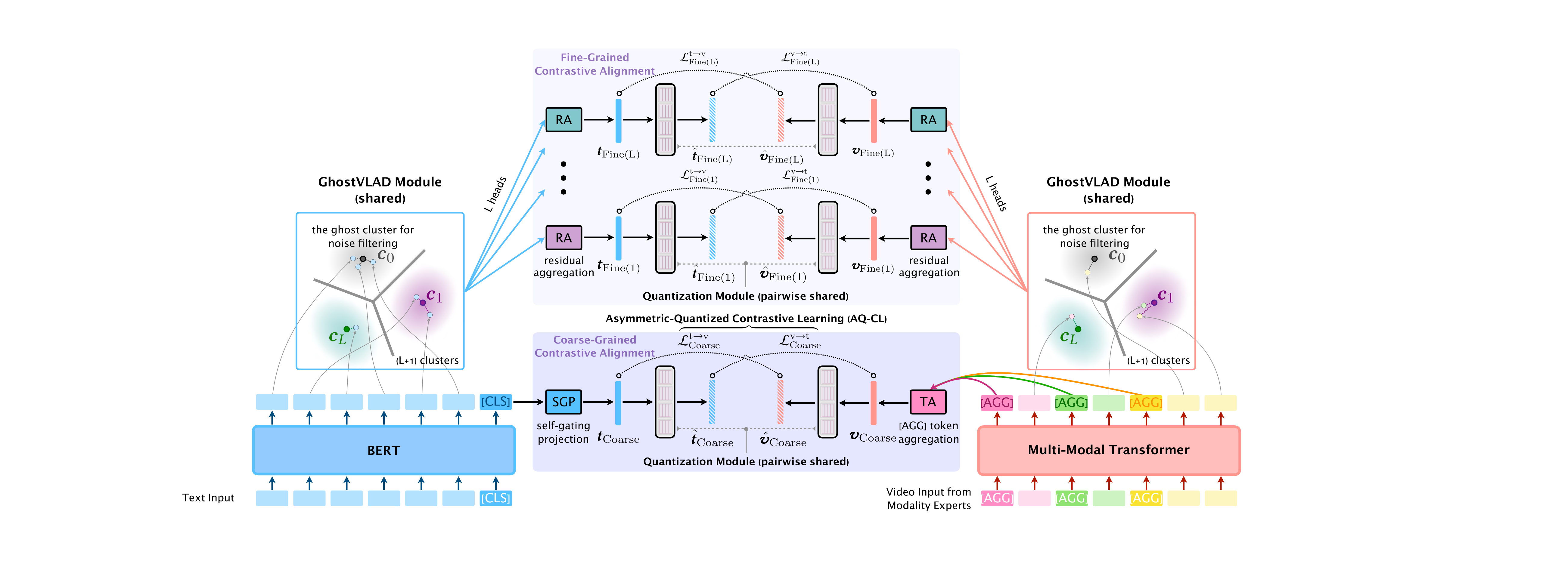}
    \caption{The framework of HCQ. 
    First, we forward the tokenized text and video features to the BERT encoder and the multi-modal video transformer, respectively. 
    Then, for the coarse-grained level, we get text embedding by refining the ``[CLS]'' token of BERT encoder with a self-gating projection layer and pool ``[AGG]'' token embeddings of the multi-modal transformer to get video embedding. 
    Next, we further process word tokens and video feature tokens with a shared GhostVLAD module~\cite{GhostVLAD} for the fine-grained levels, yielding $L$ local embeddings with latent concept information. 
    Finally, we align the text-view and video-view representations at each level by Asymmetric-Quantized Contrastive Learning (AQ-CL).
    \emph{Best viewed in color.}}
    \label{fig:arc}
\end{figure*}

\section{Method}
\label{sec:method}

\subsection{Problem Formulation and Model Overview}
\label{subsec:overview}
The goal of cross-view video retrieval (\ie ``text-video retrieval'' in this paper) is to find the matched video for each query in another view (\ie ``a text''). 
To achieve this goal, a typical model sets up independent encoders for learning texts and videos representations. 
Based on the learned representations, the model learns to maximize the measured similarities between the matched text-video pairs. 
To this end, we propose the \textbf{H}ybrid \textbf{C}ontrastive \textbf{Q}uantization (\textbf{HCQ}), which learns both coarse-grained and fine-grained quantization for efficient cross-view video retrieval. 
Given a training set $\calD=\dak{(\calT_i, \calV_i)}_{i=1}^{N_\calD}$ of $N_\calD$ matched text-video pairs, where the $i$-th pair contains a sentence $\calT_i$ and a video $\calV_i$. 
The goal of HCQ is to jointly learn view-specific quantizers $\calQ^\calT$ and $\calQ^\calV$ for texts and videos. 

As shown in Fig.\ref{fig:arc}, HCQ consists of four main components:
(\textbf{i}) A pre-trained BERT encoder~\cite{BERT}, denoted by $f(\cdot,\bmtheta^f)$, for extracting contextual word embeddings and text representation (\ie the ``[CLS]'' token).  We describe it in \S~\ref{subsec:text_enc}.
(\textbf{ii}) A multi-modal transformer~\cite{MMT}, denoted by $g(\cdot,\bmtheta^g)$, to encode video features from multiple experts, which not only captures temporal dynamics but also leverages multi-modal cues within the video view. We describe it in \S~\ref{subsec:vid_enc}.
(\textbf{iii}) A GhostVLAD module~\cite{GhostVLAD} which shared across both views, denoted by $h(\cdot,\bmtheta^h)$, for further exploring fine-grained, latent concept-aware information. We describe it in \S~\ref{subsec:hybrid_grained}. 
(\textbf{iv}) A series of $L+1$ quantization modules for learning and generating multi-level quantized representations, including one for coarse-grained level, denoted by $q_\text{Coarse}(\cdot,\bmtheta_\text{Coarse})$, and the rest $L$ for fine-grained levels, denoted by $\dak{q_\text{Fine($l$)}(\cdot,\bmtheta_\text{Fine($l$)})}_{l=1}^L$. 
We align cross-view representations by performing Asymmetric-Quantized Contrastive Learning (AQ-CL) at each level, which provides solid supervision for quantization learning. 
We depict the details in \S~\ref{subsec:aqcl}.

\subsection{Text Feature Encoding}
\label{subsec:text_enc}
With the impressive performance on language understanding, BERT \cite{BERT} has become a powerful backbone for text representation. 
Motivated by the successful practice~\cite{MMT} in text-video retrieval, we adopt the BERT encoder for text features. 
Concretely, for an input sentence $\calT$, we tokenize its $N_\calT$ words, append special tokens (\eg ``[CLS]'', and ``[SEP]''), and add position embeddings. 
After that, we forward the input sequence to the BERT encoder, $f(\cdot,\bmtheta_f)$, and compute the output embeddings. 
Let $\bme^t_\text{[CLS]}$ and $\dak{\bme^t_i}_{i=1}^{N_\calT}$ denote the output embeddings of the ``[CLS]'' token and the words in $\calT$, respectively. 
We define the whole process by 
\begin{gather}
    \bme^t_i=f(\calT)_i\in\bbR^{D^t},\,i=\text{[CLS]},\Range{N_\calT}.
\end{gather}

\subsection{Video Feature Encoding}
\label{subsec:vid_enc}
The multi-modal content and the temporal dynamics make video representation challenging. 
We make full use of multi-modal cues in HCQ.
First, for the input video $\calV$, we extract different kinds of video features (\eg appearance, motion, audio) with the help of a bunch of modality experts~\cite{CE,MMT}: $\calE=\dak{\seq{E}{N_\calE}}$. 
Then, we project all of these features to $D$-dimensional embeddings, which serve as the input embeddings of feature tokens. 
Next, we insert several modality-specific ``[AGG]'' (\ie feature aggregation) tokens to the video sequence, after which we add functional embeddings to indicate temporal and modal differences.
Finally, we forward the input sequence to the multi-modal transformer~\cite{MMT}, $g(\cdot,\bmtheta_g)$, to compute the output embeddings. 
Let $\bme^v_{\text{[AGG]}_1},\cdots,\bme^v_{\text{[AGG]}_{N_\calE}}$ denote the embeddings of modality-specific ``[AGG]'' tokens, $\dak{\bme^v_i}_{i=1}^{N_\calV}$ denote the output embeddings of all feature tokens, and $N_\calV$ denotes the total number of feature tokens \wrt $\calV$. 
Analogous to the text part, we define the whole process of video feature encoding by 
\begin{gather}
    \bme^v_i=g(\calT)_i\in\bbR^D,\,i=\text{[AGG]}_1,\cdots,\text{[AGG]}_{N_\calE},\Range{N_\calV}.
\end{gather}

\subsection{Hybrid-grained Representation}
\label{subsec:hybrid_grained}
HCQ learns both coarse-grained and fine-grained representations, which provide complementary understandings for texts and videos and preserve comprehensive semantic information. 

\subsubsection{Coarse-grained Representation}
The coarse-grained representation condenses global information, providing a direct manner for cross-view matching.
For the text view, we adopt a projection layer, $\phi:\bbR^{D^t}\mapsto\bbR^D$, with a self-gating mechanism~\cite{MoEE,MMT}. It serves as a multi-modality-aware refinement on the ``[CLS]'' embedding.
We compute the coarse-grained text embedding by
\begin{equation}
    \bmt_\text{Coarse}=\phi(\bme^t_\text{[CLS]})=\sum_{i=1}^{N_\calE}\omega_i(\bme^t_\text{[CLS]})\cdot\psi_i(\bme^t_\text{[CLS]}),
\end{equation}
where $\psi_i:\bbR^{D^t}\mapsto\bbR^D$ is a linear transformation with $\ell_2$ normalization, corresponding to the modality expert $E_i$. We follow \citet{MoEE} to obtain the gating weights by a linear layer and a subsequent softmax operation, namely,
\begin{equation}
\omega_i(\bme^t_\text{[CLS]})=\frac{\exp(\bmh_i^\top\bme^t_\text{[CLS]})}{\sum_{j=1}^{N_\calE}\exp(\bmh_j^\top\bme^t_\text{[CLS]})},\,i=\Range{N_\calE}.
\end{equation}
For the video view, we use a simple average pooling to aggregate multiple ``[AGG]'' tokens, namely, 
\begin{equation}
    \bar{\bme}^v_\text{[AGG]}=\frac1{N_\calE}\sum_{i=1}^{N_\calE}\bme^v_{\text{[AGG]}_i}.
\end{equation}
We then apply $\ell_2$ normalization to $\bar{\bme}^v_\text{[AGG]}$ and obtain $\bmv_\text{Coarse}$.
Empirically, we find it better than using a ``[CLS]'' token for video.

\subsubsection{Fine-grained Representation}
The fine-grained representation preserves concept-aware semantic information, enabling more precise matching across views. 
To this end, we learn a structural space with a GhostVLAD \cite{GhostVLAD} module. 
Specifically, we learn $L+1$ $D$-dimensional cluster centroids, $\dak{\bmc_0,\seq{\bmc}{L}}$, in a shared embedding space. 
We designate $\bmc_0$ as the ``ghost'' centroid for filtering noise, \eg uninformative words in a sentence, and background features for a video. 
To obtain fine-grained text embeddings, we first linearly project the $D^t$-dimensional word token embeddings, $\dak{\bme^t_i}_{i=1}^{N_\calT}$, to the $D$-dimensional space. 
Then, we forward the projected embeddings, $\dak{\tilde{\bme}^t_i}_{i=1}^{N_\calT}$, to the GhostVLAD. 
Each word embedding will be softly assigned to all clusters. 
For instance, the assignment score of $\tilde{\bme}^t_i$ \wrt the $l$-th cluster is defined as
\begin{equation}
    a^t_{i,l}=\frac{\exp(\text{BatchNorm}(\bmw_l^\top\tilde{\bme}^t_i))}{\sum_{l'=0}^L\exp(\text{BatchNorm}(\bmw_{l'}^\top\tilde{\bme}^t_i))},
\end{equation}
where $\text{BatchNorm}(\cdot)$ denotes the batch normalization~\cite{BN} operation,  $\bmW=[\bmw_0,\seq{\bmw}{L}]$ is the matrix of trainable parameters. 
After clustering $\dak{\tilde{\bme}^t_i}_{i=1}^{N_\calT}$, we aggregate the residual embeddings of each cluster, except the ``ghost'' cluster.
Take the $l$-th cluster as an example, the aggregated residual embedding is computed by
\begin{equation}
    \bmr^t_l=\sum_{i=1}^{N_\calT}a^t_{i,l}\cdot(\tilde{\bme}^t_i-\bmc_l)
\end{equation}
We obtain the fine-grained text embeddings, $\{\bmt_{\text{Fine}(l)}\}_{l=1}^L$, by performing $\ell_2$ normalization on $\{\bmr^t_l\}_{l=1}^L$. 
To obtain fine-grained video embeddings, the process is analogous except for the initial projection from $\bbR^{D^t}$ to $\bbR^D$. 
We directly forward $\dak{\bme^v_i}_{i=1}^{N_\calV}$ to the GhostVLAD and obtain $\{\bmv_{\text{Fine}(l)}\}_{l=1}^L$.

\subsection{Trainable Quantization Modules}
\label{subsec:trainable}
After obtaining coarse-grained embeddings and multi-level fine-grained embeddings of texts and videos, we consider applying quantization to them.
At each level, there is one quantization module shared across text and video. 
Note that those widely adopted quantization methods~\cite{PQ,OPQ} in practical Web search engines are hard to be integrated into the deep learning framework.
The reason is that the codeword assignment step is clustering-based and can not be trained by back-propagation. 
To jointly learn quantization, we adopt a trainable scheme in our quantization modules. 

Without loss of generality and for a concise interpretation, this section describes the quantization module of a specific level with no subscript like ``$_\text{Coarse}$'' or ``$_{\text{Fine}(l)}$''. 
We denote the quantization codebooks of this quantization module by $\calC=\bmC^1\times\bmC^2\times\cdots\times\bmC^M$, where the $m$-th codebook $\bmC^m\in\bbR^{K\times d}$ consists of $K$ codewords $\seq{\bmc^m}{K}\in\bbR^d$. 
Suppose that $\bbR^D\equiv\bbR^{Md}$ (where $M$ and $d$ are both positive integers)
Suppose that $\bbR^D\equiv\bbR^{Md}$ (where $M$ and $d$ are both positive integers), $\bmt,\bmv\in\bbR^D$ are text embedding and video embedding at the same representation level. 
In this case, $\bmt$ and $\bmv$ can be divided into $M$ equal-length $d$-dimensional segments, \ie $\bmt\in\bbR^D\equiv[\bmt^1,\cdots,\bmt^M]$ and $\bmv\in\bbR^D\equiv[\bmv^1,\cdots,\bmv^M]$, where $\bmt^m,\bmv^m\in\bbR^d$, $d=D/M$, $1\le m\le M$.
Given a vector, each codebook quantize one segment respectively.
Take the text embedding vector as an example, in the $m$-th $d$-dimensional subspace, we first normalize its the segment and codewords:
\begin{equation}
    \bmt^m\leftarrow\bmt^m/\shuk{\bmt^m}_2,\ \bmc_i^m\leftarrow\bmc_i^m/\shuk{\bmc_i^m}_2.
\end{equation}
Then each segment is quantized with codebook attention by
\begin{equation}\label{equ:code_attention}
    \hat{\bmt}^m=\text{Attention}(\bmt^m,\bmC^m,\bmC^m)
    =\sum_{i=1}^Kp_i^m\bmc_i^m,
\end{equation}
where attention score $p_i^m$ is computed with softmax operation:
\begin{equation}\label{equ:alpha_softmax}
    p^m_i
    =\frac{\exp\xiaok{\alpha\cdot{\bmt^m}^\top\bmc^m_i}}{\sum_{j=1}^K\exp\xiaok{\alpha\cdot{\bmt^m}^\top\bmc^m_j}}.
\end{equation}
The softmax is a differentiable alternative to the argmax that relaxes the discrete optimization of hard codeword assignment to a trainable form. 
Finally, we get \emph{soft} quantization code 
and the \emph{soft} quantized reconstruction of $\bmt$ as
\begin{gather}
    \bmp=\text{concatenate}\xiaok{\bmp^1,\bmp^2,\cdots,\bmp^M}\in\bbR^{KM},\\
    \label{equ:concat}\hat{\bmt}= \text{concatenate}\xiaok{\hat{\bmt}^1,\hat{\bmt}^2,\cdots,\hat{\bmt}^M}.
\end{gather}
The quantization for video embedding $\bmv$ is analogous.

\subsection{Hybrid Contrastive Learning}
\label{subsec:aqcl}
To learn high-quality representations, we perform \emph{Asymmetric-Quantized Contrastive Learning (AQ-CL)} at the coarse-grained level and multiple fine-grained levels. 
Suppose that there is a batch of training pairs, $\calB=\dak{(\calT_i,\calV_i)}_{i=1}^{N_\calB}$, and we have extracted coarse-grained and fine-grained embeddings for them. 

At the coarse-grained level, we first take the $i$-th raw text embedding $\bmt_{\text{Coarse},i}$ as the \emph{query}, and the quantized video embeddings, $\dak{\hat{\bmv}_{\text{Coarse},j}}_{j=1}^{N_\calB}$, as the \emph{keys} in the contrastive learning loss~\cite{moco,simclr}. 
Then coarse-grained text-to-video AQ-CL loss \wrt $\calB$ is defined as
\begin{equation}\label{equ:coarse_t2v}
    \calL_\text{Coarse}^{t\rightarrow v}=-\frac1{N_\calB}\sum_{i=1}^{N_\calB}\log\frac{\exp(s_{\text{Coarse},ii}^{t\rightarrow v}/\tau)}{\sum_{j=1}^{N_\calB}\exp(s_{\text{Coarse},ij}^{t\rightarrow v}/\tau)},
\end{equation}
where $\tau$ is the temperature factor for contrastive learning. The text-to-video coarse-grained similarity is defined by
\begin{equation}\label{equ:t2v_sim}
    s_{\text{Coarse},ij}^{t\rightarrow v}=\bmt_{\text{Coarse},i}^\top\hat{\bmv}_{\text{Coarse},j}.
\end{equation}
Symmetrically, we take the raw video embedding $\bmv_{\text{Coarse},i}$ as the query and leave $\dak{\hat{\bmt}_{\text{Coarse},j}}_{j=1}^{N_\calB}$ as the keys. 
The coarse-grained video-to-text AQ-CL loss \wrt $\calB$ is defined as
\begin{equation}\label{equ:coarse_v2t}
    \calL_\text{Coarse}^{v\rightarrow t}=-\frac1{N_\calB}\sum_{i=1}^{N_\calB}\log\frac{\exp(s_{\text{Coarse},ii}^{v\rightarrow t}/\tau)}{\sum_{j=1}^{N_\calB}\exp(s_{\text{Coarse},ij}^{v\rightarrow t}/\tau)},
\end{equation}
where the video-to-text coarse-grained similarity is defined by
\begin{equation}\label{equ:v2t_sim}
    s_{\text{Coarse},ij}^{v\rightarrow t}=\bmv_{\text{Coarse},i}^\top\hat{\bmt}_{\text{Coarse},j}.
\end{equation}
The loss for coarse-grained representations is defined as
\begin{equation}\label{equ:coarse_CL_loss}
    \calL_\text{Coarse}=\frac12\xiaok{\calL_\text{Coarse}^{t\rightarrow v}+\calL_\text{Coarse}^{v\rightarrow t}}.
\end{equation}

The contrastive learning processes on fine-grained levels are similar to Eq.(\ref{equ:coarse_t2v})-(\ref{equ:coarse_CL_loss}). 
We implement the hybrid contrastive learning of HCQ in a multi-task manner. The hybrid loss is defined by
\begin{equation}
    \calL_\text{HCQ}=\calL_\text{Coarse}+\frac1L\sum_{l=1}^L\calL_{\text{Fine}(l)},
\end{equation}
where $\calL_{\text{Fine}(l)}$ is the AQ-CL loss for the $l$-th fine-grained level.

\subsection{Indexing and Retrieval}
\label{subsec:indx_and_retr}
In this section, we take the text-to-video retrieval as an example to describe the quantization-based inference. 

First, we encode the database (\ie the videos) with hard quantization. 
For the coarse-grained representations $\bmv_\text{Coarse}$, we divide it into $M$ equal-length segments and perform hard quantization by
\begin{gather}
    i^m=\argmax{1\le i\le K}{{\bmv^m_\text{Coarse}}^\top\bmc_i^m},\quad
    \hat{\bmv}^m_\text{Coarse}=\bmc_{i^m}^m,\quad1\le m\le M.
\end{gather} 

Then, given the coarse-grained embedding of a retrieval query (\ie a sentence), we cut the vector into $M$ equal-length segments, \ie $\bmt_\text{Coarse}=[\bmt_\text{Coarse}^1; \bmt_\text{Coarse}^2; \cdots; \bmt_\text{Coarse}^M]$, $\bmt_\text{Coarse}^m\in\bbR^d$. 
To obtain coarse-grained text-video matching scores, we adopt Asymmetric Quantized Similarity (AQS)~\cite{PQ} as the metric, which computes the inner product between $\bmt_\text{Coarse}$ and the coarse-grained quantized embedding of the $x$-th database item $\hat{\bmv}_{\text{Coarse},x}$ by
\begin{equation}
    \text{AQS}_\text{Coarse}^{t\rightarrow v}(\calT,\calV_x)=\sum_{m=1}^M{\bmt_\text{Coarse}^m}^\top\hat{\bmv}_{\text{Coarse},x}^m=\sum_{m=1}^M{\bmt_\text{Coarse}^m}^\top\bmc_{i^m_x}^m.
\end{equation}

To accelerate the computation, we can set up a query-specific lookup table $\bmXi_\text{Coarse}\in\bbR^{M\times K}$ for each $\bmt_\text{Coarse}$, which stores the pre-computed similarities between the segments of $\bmt_\text{Coarse}$ and all codewords. 
Specifically, $\Xi^m_{\text{Coarse},i^m}={\bmt_\text{Coarse}^m}^\top\bmc_{i^m}^m$. 
Hence, the AQS can be efficiently computed by summing chosen items from the lookup table according to the quantization code, namely,
\begin{equation}
    \text{AQS}_\text{Coarse}^{t\rightarrow v}(\calT,\calV_x)=\sum_{m=1}^M\Xi^m_{\text{Coarse},i^m_x},
\end{equation}
where $i^m_x$ is the index of the assigned codeword in the $m$-th codebook. 
The AQS computations at fine-grained levels are analogous. 

Finally, we define the hybrid-grained text-video similarity by
\begin{equation*}
    s^{t\rightarrow v}(\calT,\calV_x)=\text{AQS}_\text{Coarse}^{t\rightarrow v}(\calT,\calV_x)+\frac1L\sum_{l=1}^L\text{AQS}_{\text{Fine}(l)}^{t\rightarrow v}(\calT,\calV_x),
\end{equation*}
which is used to rank and retrieve videos.

\section{Experiments}
\label{sec:experiments}

\subsection{Research Questions}
\label{subsec:rq}
We evaluate the proposed method by conducting experiments on three datasets. We aim to answer the following research
questions:\setlist{nolistsep}
\begin{itemize}
	\item[\textbf{RQ1:}] Compared with state-of-the-art cross-view (text-video) retrieval models, can HCQ keep competitive performance while showing high efficiency in computation and stoarge?
	\item[\textbf{RQ2:}] On the specific task of cross-view video retrieval, does HCQ outperform other embedding compression approaches?
	\item[\textbf{RQ3:}] How do different strategies contribute to the effectiveness of HCQ?
\end{itemize}

\subsection{Experimental Setup}
\label{subsec:setup}
\subsubsection{Datasets}
We conduct experiments on three benchmark datasets. 
(\textbf{i}) \textbf{MSRVTT}~\cite{MSRVTT} dataset contains 10,000 YouTube videos collected from a commercial Web search engine. 
Each video is annotated with 20 captions. 
We experiment with three data splits. 
The first two splits, denoted by ``1k-A'' and ``1k-B'', are from \citet{JSFusion} and \citet{MoEE}, respectively. 
Each of them selects 1,000 videos for testing and the remaining videos are used for training. 
The last one is the official split, which uses 6,513 videos for training, 497 videos for validation and the remaining 2,990 videos for testing. 
Unless otherwise specified, the MSRVTT results are with “1k-A”.
(\textbf{ii}) \textbf{LSMDC}~\cite{LSMDC} dataset contains 118,081 video clips from 202 movies. 
Each of video clips is annotated with a caption. 
We follow~\citet{MMT} to split a evaluation set of 1,000 video clips, and the remaining video clips are for training. 
(\textbf{iii}) \textbf{ActivityNet Captions}~\cite{ActivityNet} dataset contains 20,000 YouTube videos, each of which is paired with several captions. 
Following~\citet{HSE,MMT}, we train the model with 10,009 videos and perform evaluations on the ``val1'' split, which includes 4,917 videos. 

\subsubsection{Metrics}
We evaluate the models from different aspects. 
On the performance, we adopt standard metrics, including \textbf{recall at rank $N$} (\textbf{R@$N$}, higher is better) and \textbf{median rank} (\textbf{MdR}, lower is better). 
On the efficiency, we adopt the \textbf{average query time} for one query (lower is better) for the computation efficiency. We adopt the \textbf{storage size} of representations (lower is better) to evaluate storage (or memory) efficiency.

\subsubsection{Baselines}
First, we compare our model with 21 classic or state-of-the-art methods in text-video retrieval:
\setlist{nolistsep}
\begin{enumerate}[leftmargin=2em]
    \item Continous representation model:
    \begin{enumerate}[label=(\textbf{\roman*}),leftmargin=0.5em]
        \item Adopt basic sequential modeling: \textbf{VSE}~\cite{VSE}, \textbf{VSE++}~\cite{VSE2}, \textbf{HT}~\cite{HowTo100M}, \textbf{JPose}~\cite{JPose}, \textbf{W2VV}~\cite{W2VV}, \textbf{Dual Encoding}~\cite{DualEncoding}, \textbf{E2E}~\cite{E2E}, \textbf{CT-SAN}~\cite{CTSAN}, \textbf{FSE}~\cite{HSE}, and \textbf{HSE}~\cite{HSE};
        \item Graph-based and tree-based models:  \textbf{HGR}~\cite{HGR}, \textbf{ViSERN}~\cite{ViSENR}, \textbf{HCGC}~\cite{HCGC}, \textbf{HANet}~\cite{HANet}, and \textbf{TCE}~\cite{TreeAug};
        \item Leverage multi-modal cues (e.g., audio, speech, OCR, and ASR information):  \textbf{JSFusion}~\cite{JSFusion}, \textbf{\citet{mithun2018learning}}, \textbf{MoEE}~\cite{MoEE}, \textbf{CE}~\cite{CE}, and \textbf{MMT}~\cite{MMT};
    \end{enumerate}
    \item Hashing-based representation model:  \textbf{S$^2$Bin}~\cite{qi2021semantics}. 
\end{enumerate}

From another perspective, we also compare our model with hashing methods, including:
\begin{enumerate}[label=(\textbf{\roman*}),leftmargin=2em]
\item Commonly-adopted post-processing hashing methods: \textbf{LSH}~\cite{LSH}, \textbf{PQ}~\cite{PQ}, and \textbf{OPQ}~\cite{OPQ};
\item Deep cross-modal image hashing methods: 
\textbf{DCMH}~\cite{DCMH};
\item Recently-proposed dense retrieval method on jointly learning query encoder and item quantization: \textbf{JPQ}~\cite{JPQ}. 
\end{enumerate}
Note that DCMH and JPQ only learn global  (\ie coarse-grained) representations for images or texts. 
In the light of their original paradigm, we implement them with coarse-grained embeddings. 

\subsubsection{Implementation Details}
We use the multi-modal features provided by \citet{MMT} as the input of the video transformer. 
These features were extracted by pre-trained experts on motion, appearance, audio, speech, scene, face and OCR. 
On MSRVTT, we leverage all modality features. 
On LSMDC dataset, we do not use speech features due to their absence in the release~\cite{MMT}. 
On ActivityNet Captions dataset, we use motion and audio features since the release only contains them. 
The multi-modal video transformer is composed of 4 layers and 4 heads. 
The dropout rate of video transformer is set to 0.1 and the embedding dimension $D$ is set to 512.  
For the shared GhostVLAD module, the active cluster (\ie non-ghost cluster) number $L$ is set to 7 by default. 
We set one quantization module for each of the coarse-grained level and multiple fine-grained levels, resulting in $L$ quantization modules. 
For each quantization module, the codeword number of each codebook, $K$, is set to 256 such that each embedding is encoded by $M\log_2K=8M$ bits (\ie $M$ bytes).
The number of sub-codebook in each quantization module, $M$, is set to 32 by default. 
Thus the default code-length of each quantization module is 256 bits (\ie 32 bytes), and the default code-length of HCQ is $256L$ bits (\ie $32L$ bytes) in total.
The scaling factor for quantization, $\alpha$, is set to 1. 
For the contrastive learning, we set the temperature factor, $\tau$, to $0.05$ on MSRVTT and LSMDC datasets and set it to $0.07$ on ActivityNet Captions dataset by default. 
We use the Adam optimizer~\cite{Adam} with an initial learning rate of 5e-5, decaying every 1K steps. 
The training batchsize is set to $128$ on MSRVTT and LSMDC datasets and is set to $32$ on ActivityNet Captions dataset. 
We use the ``BERT-base-cased'' to initialize the text encoder and finetune it with a dropout rate of 0.1. 
For more details, please see the Github repository\footnote{Code and configurations are available at \url{https://github.com/gimpong/WWW22-HCQ}.}.

\begin{table}[t]
\caption{Retrieval performance on MSRVTT dataset~\cite{MSRVTT}. ``Text $\rightarrow$ Video'' and ``Video $\rightarrow$ Text'' denote the text-to-video retrieval and video-to-text retrieval, respectively.}
\centering
\resizebox{\columnwidth}{!}
{
\setlength{\tabcolsep}{0.1em}{
    \begin{tabular}{llccccccccccc}
    \toprule
     &  &  &  & \multicolumn{4}{c}{Text $\rightarrow$   Video} &  & \multicolumn{4}{c}{Video $\rightarrow$   Text} \\
    \cmidrule{5-8} \cmidrule{10-13}
    \multirow{-2}{*}{Method} &  & \multirow{-2}{*}{Split} &  & R@1$\uparrow$ & R@5$\uparrow$ & R@10$\uparrow$ & MdR$\downarrow$ &  & R@1$\uparrow$ & R@5$\uparrow$ & R@10$\uparrow$ & MdR$\downarrow$ \\
    \toprule
    JSFusion~\cite{JSFusion} &  & 1k-A &  & 10.2 & 31.2 & 43.2 & 13 &  & - & - & - & - \\
    HT~\cite{HowTo100M} &  & 1k-A &  & 14.9 & 40.2 & 52.8 & 9 &  & - & - & - & - \\
    MoEE~\cite{MoEE} &  & 1k-A &  & 16.8 & 41.0 & 54.4 & 9 &  & - & - & - & - \\
    CE~\cite{CE} &  & 1k-A &  & 20.9 & 48.8 & 62.4 & 6 &  & 20.6 & 50.3 & 64.0 & 5.3 \\
    MMT~\cite{MMT} &  & 1k-A &  & 24.6 & 54.0 & 67.1 & \textbf{4} &  & 24.4 & 56.0 & 67.8 & \textbf{4} \\
    TCE~\cite{TreeAug} &  & 1k-A &  & 17.1 & 39.9 & 53.7 & 9 &  & - & - & - & - \\
    \rowcolor[HTML]{EFFBEC} 
    HCQ   (Ours) &  & 1k-A &  & \textbf{25.9} & \textbf{54.8} & \textbf{69.0} & 5 & \textbf{} & \textbf{26.3} & \textbf{57.0} & \textbf{70.1} & \textbf{4} \\ 
    \midrule
    MoEE~\cite{MoEE} &  & 1k-B &  & 13.6 & 37.9 & 51.0 & 10 &  & - & - & - & - \\
    JPose~\cite{JPose} &  & 1k-B &  & 14.3 & 38.1 & 53.0 & 9 &  & 16.4 & 41.3 & 54.4 & 8.7 \\
    MoEE-COCO~\cite{MoEE} &  & 1k-B &  & 14.2 & 39.2 & 53.8 & 9 &  & - & - & - & - \\
    CE~\cite{CE} &  & 1k-B &  & 18.2 & 46.0 & 60.7 & 7 &  & 18.0 & 46.0 & 60.3 & 6.5 \\
    MMT~\cite{MMT} &  & 1k-B &  & 20.3 & 49.1 & 63.9 & 6 &  & 21.1 & 49.4 & 63.2 & 6 \\
    TCE~\cite{TreeAug} &  & 1k-B &  & 17.1 & 39.9 & 53.7 & 9 &  & - & - & - & - \\
    \rowcolor[HTML]{EFFBEC} 
    HCQ   (Ours) &  & 1k-B &  & \textbf{22.5} & \textbf{51.5} & \textbf{65.9} & \textbf{5} & \textbf{} & \textbf{23.7} & \textbf{52.2} & \textbf{66.9} & \textbf{5} \\
    \midrule
    VSE~\cite{VSE} &  & Full &  & 5.0 & 16.4 & 24.6 & 47 &  & 7.7 & 20.3 & 31.2 & 28 \\
    VSE++~\cite{VSE2} &  & Full &  & 5.7 & 17.1 & 24.8 & 65 &  & 10.2 & 25.4 & 35.1 & 25 \\
    \citet{mithun2018learning} &  & Full &  & 7.0 & 20.9 & 29.7 & 38 &  & 12.5 & 32.1 & 42.4 & 16 \\
    W2VV~\cite{W2VV} &  & Full &  & 6.1 & 18.7 & 27.5 & 45 &  & 11.8 & 28.9 & 39.1 & 21 \\
    Dual   Encoding~\cite{DualEncoding} &  & Full &  & 7.7 & 22.0 & 31.8 & 32 &  & 13.0 & 30.8 & 43.3 & 15 \\
    TCE~\cite{TreeAug} &  & Full &  & 7.7 & 22.5 & 32.1 & 30 &  & - & - & - & - \\
    HGR~\cite{HGR} &  & Full &  & 9.2 & 26.2 & 36.5 & 24 &  & 15.0 & 36.7 & 48.8 & 11 \\
    E2E~\cite{E2E} &  & Full &  & 9.9 & 24.0 & 32.4 & 29.5 &  & - & - & - & - \\
    CE~\cite{CE} &  & Full &  & 10.0 & 29.0 & 41.2 & 16 &  & 15.6 & 40.9 & 55.2 & 8.3 \\
    HANet~\cite{HANet} &  & Full &  & 9.3 & 27.0 & 38.1 & 20 &  & 16.1 & 39.2 & 52.1 & 9 \\
    ViSERN~\cite{ViSENR} &  & Full &  & 7.9 & 23.0 & 32.6 & 30 &  & 13.1 & 30.1 & 43.5 & 15 \\
    HCGC~\cite{HCGC} &  & Full &  & 9.7 & 28.0 & 39.2 & 19 &  & 17.1 & 40.5 & 53.2 & 9 \\
    S$^2$Bin~\cite{qi2021semantics} &  & Full &  & 7.9 & 22.5 & 32.2 & 31 &  & 13.3 & 32.5 & 43.7 & 15 \\
    \rowcolor[HTML]{EFFBEC} 
    HCQ   (Ours) &  & Full &  & \textbf{15.2} & \textbf{38.5} & \textbf{51.0} & \textbf{10} & \textbf{} & \textbf{18.3} & \textbf{44.9} & \textbf{59.1} & \textbf{7} \\
    \bottomrule
    \end{tabular}
}
}
\label{tab:MSRVTT}
\end{table}

\begin{table}[t]
\caption{Retrieval performance on LSMDC dataset~\cite{LSMDC}. ``Text $\rightarrow$ Video'' and ``Video $\rightarrow$ Text'' denote the text-to-video retrieval and video-to-text retrieval, respectively.}
\centering
\resizebox{\columnwidth}{!}
{
\setlength{\tabcolsep}{0.2em}{
\begin{tabular}{llccccccccc}
\toprule
 &  & \multicolumn{4}{c}{Text $\rightarrow$   Video} &  & \multicolumn{4}{c}{Video $\rightarrow$   Text} \\
\cmidrule{3-6} \cmidrule{8-11}
\multirow{-2}{*}{Method} &  & R@1$\uparrow$ & R@5$\uparrow$ & R@10$\uparrow$ & MdR$\downarrow$ &  & R@1$\uparrow$ & R@5$\uparrow$ & R@10$\uparrow$ & MdR$\downarrow$ \\
\midrule
CT-SAN   ~\cite{CTSAN} &  & 5.1 & 16.3 & 25.2 & 46 &  & - & - & - & - \\
JSFusion   ~\cite{JSFusion} &  & 9.1 & 21.2 & 34.1 & 36 &  & - & - & - & - \\
CCA   ~\cite{CCA} &  & 7.5 & 21.7 & 31 & 33 &  & - & - & - & - \\
MoEE~\cite{MoEE} &  & 9.3 & 25.1 & 33.4 & 27 &  & - & - & - & - \\
MoEE-COCO   ~\cite{MoEE} &  & 10.1 & 25.6 & 34.6 & 27 &  & - & - & - & - \\
CE   ~\cite{CE} &  & 11.2 & 26.9 & 34.8 & 25.3 &  & - & - & - & - \\
MMT~\cite{MMT} &  & 13.2 & 29.2 & 38.8 & 21 &  & 12.1 & 29.3 & 37.9 & 22.5 \\
TCE~\cite{TreeAug} &  & 10.6 & 25.8 & 35.1 & 29 &  & - & - & - & - \\
\rowcolor[HTML]{EFFBEC} 
HCQ   (Ours) &  & \textbf{14.5} & \textbf{33.6} & \textbf{43.1} & \textbf{18.5} &  & \textbf{13.7} & \textbf{33.2} & \textbf{42.8} & \textbf{17} \\
\bottomrule
\end{tabular}}
}
\label{tab:LSMDC}
\end{table}

\begin{table}[t]
\caption{Retrieval performance on ActivityNet Captions dataset~\cite{ActivityNet}. ``Text $\rightarrow$ Video'' and ``Video $\rightarrow$ Text'' denote the text-to-video retrieval and video-to-text retrieval.}
\centering
\resizebox{\columnwidth}{!}
{
\setlength{\tabcolsep}{0.2em}{
\begin{tabular}{llccccccccc}
\toprule
 &  & \multicolumn{4}{c}{Text $\rightarrow$   Video} &  & \multicolumn{4}{c}{Video $\rightarrow$   Text} \\
\cmidrule{3-6} \cmidrule{8-11}
\multirow{-2}{*}{Method} &  & R@1$\uparrow$ & R@5$\uparrow$ & R@50$\uparrow$ & MdR$\downarrow$ &  & R@1$\uparrow$ & R@5$\uparrow$ & R@50$\uparrow$ & MdR$\downarrow$ \\
\midrule
FSE~\cite{HSE} &  & 18.2 & 44.8 & 89.1 & 7 &  & 16.7 & 43.1 & 88.4 & 7 \\
HSE~\cite{HSE} &  & 20.5 & 49.3 & - & - &  & 18.7 & 48.1 & - & - \\
CE~\cite{CE} &  & 18.2 & 47.7 & 91.4 & 6 &  & 17.7 & 46.6 & 90.9 & 6 \\
MMT~\cite{MMT} &  & \textbf{22.7} & \textbf{54.2} & \textbf{93.2} & \textbf{5} &  & 22.9 & 54.8 & \textbf{93.1} & \textbf{4.3} \\
\rowcolor[HTML]{EFFBEC} 
HCQ   (Ours) &  & 22.2 & 53.7 & 91.2 & \textbf{5} &  & \textbf{23.0} & \textbf{54.9} & 91.4 & 5 \\ 
\bottomrule
\end{tabular}
}}
\label{tab:ActNet}
\end{table}

\begin{figure}[t]
    \centering
    \includegraphics[width=\columnwidth]{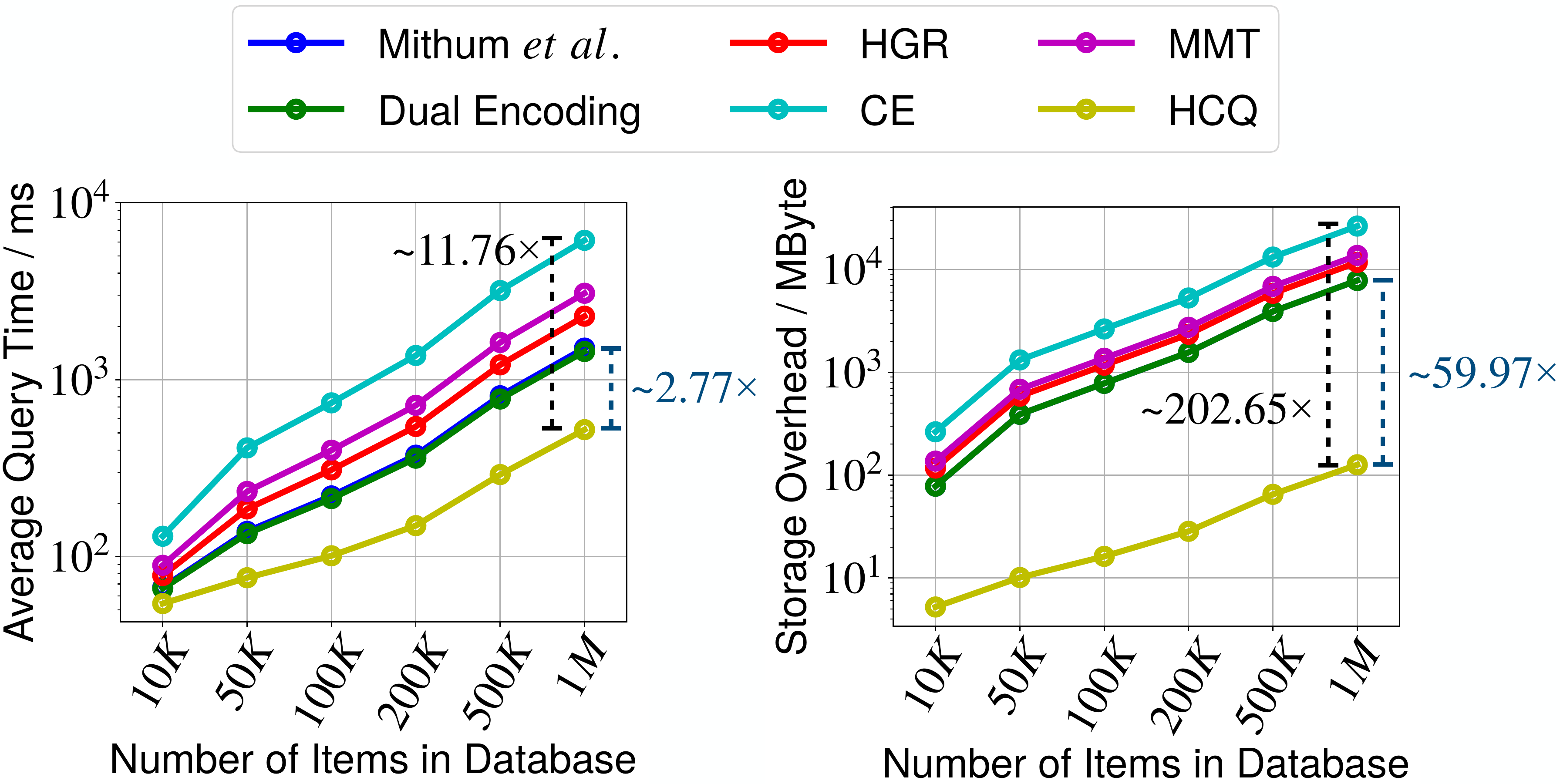}
    \caption{The average query time and storage (memory) overhead on the task of text-to-video retrieval.}
    \label{fig:efficiency}
\end{figure}

\begin{table*}[t]
\caption{The text-to-video retrieval performance comparison with hashing approaches.
For post-processing baselines, we remove the quantization modules of HCQ and train the remaining part (denoted by HCT) by raw embedding-based hybrid contrastive learning. 
Then we apply these methods to the learned embeddings and generate compressed representations. 
For joint-learning baselines, we further remove the fine-grained representation modules from HCT and use the remain part (denoted by DT) as the backbone, which keeps consistent with their coarse-grained representation manners.}
\centering
\resizebox{\textwidth}{!}
{
\setlength{\tabcolsep}{0.25em}{
    \begin{tabular}{llcccccccccccccc}
    \toprule
     & \phantom{00000} & \multicolumn{4}{c}{MSRVTT} &  & \multicolumn{4}{c}{LSMDC} &  & \multicolumn{4}{c}{ActivityNet Captions} \\
    \cmidrule{3-6}
    \cmidrule{8-11}
    \cmidrule{13-16}
    \multirow{-2}{*}{Method} &  & R@1$\uparrow$ & R@5$\uparrow$ & R@10$\uparrow$ & MdR$\downarrow$ &  & R@1$\uparrow$ & R@5$\uparrow$ & R@10$\uparrow$ & MdR$\downarrow$ &  & R@1$\uparrow$ & R@5$\uparrow$ & R@50$\uparrow$ & MdR$\downarrow$ \\
    \toprule
    \cellcolor[HTML]{E9EAFF}HCT & \cellcolor[HTML]{E9EAFF}\phantom{00000} & \cellcolor[HTML]{E9EAFF}27.8 & \cellcolor[HTML]{E9EAFF}58.9 & \cellcolor[HTML]{E9EAFF}70.0 & \cellcolor[HTML]{E9EAFF}4 & \cellcolor[HTML]{E9EAFF}\phantom{00000} & \cellcolor[HTML]{E9EAFF}16.4 & \cellcolor[HTML]{E9EAFF}34.1 & \cellcolor[HTML]{E9EAFF}43.1 & \cellcolor[HTML]{E9EAFF}17 & \cellcolor[HTML]{E9EAFF}\phantom{00000} & \cellcolor[HTML]{E9EAFF}23.1 & \cellcolor[HTML]{E9EAFF}55.0 & \cellcolor[HTML]{E9EAFF}92.6 & \cellcolor[HTML]{E9EAFF}5 \\
    HCT+LSH~\cite{LSH} & \phantom{00000} & 22.5 & 51.2 & 63.4 & 5 & \phantom{00000} & 12.6 & 28.8 & 36.7 & 22 & \phantom{00000} & 21.0 & 53.1 &  90.8 & 5 \\
    HCT+PQ~\cite{PQ} & \phantom{00000} & 22.8 & 52.3 & 66.0 & 5 &  & 13.5 & 29.5 & 38.9 & 20 & \phantom{00000} & 19.8 & 50.4 & 91.0 & 5 \\
    HCT+OPQ~\cite{OPQ} & \phantom{00000} & 23.5 & 53.6 & 67.6 & 5 & \phantom{00000} & 13.7 & 29.4 & 40.2 & 19 & \phantom{00000} & 20.4 & 52.5 & 91.2 & 5 \\
    DT+DCMH~\cite{DCMH} & \phantom{00000} & 19.0 & 48.4 & 62.2 & 6 & \phantom{00000} & 10.0 & 25.8 & 36.0 & 22 & \phantom{00000} & 12.3 & 38.3 & 84.6 & 8.5 \\
    DT+JPQ~\cite{JPQ} & \phantom{00000} & 18.9 & 46.8 & 60.8 & 6 & \phantom{00000} & 9.5 & 23.4 & 34.3 & 25 & \phantom{00000} & 17.1 & 46.4 & 90.1 & 6 \\ \cellcolor[HTML]{EFFBEC}HCQ (Ours) & \cellcolor[HTML]{EFFBEC}\phantom{00000} & \cellcolor[HTML]{EFFBEC}\textbf{25.9} & \cellcolor[HTML]{EFFBEC}\textbf{54.8} & \cellcolor[HTML]{EFFBEC}\textbf{69.0} & \cellcolor[HTML]{EFFBEC}\textbf{5} &
    \cellcolor[HTML]{EFFBEC}\phantom{00000} & \cellcolor[HTML]{EFFBEC}\textbf{14.5} & \cellcolor[HTML]{EFFBEC}\textbf{33.6} & \cellcolor[HTML]{EFFBEC}\textbf{43.1} & \cellcolor[HTML]{EFFBEC}\textbf{18.5} & \cellcolor[HTML]{EFFBEC}\phantom{00000} & \cellcolor[HTML]{EFFBEC}\textbf{22.2} & \cellcolor[HTML]{EFFBEC}\textbf{53.7} & \cellcolor[HTML]{EFFBEC}\textbf{91.2} & \cellcolor[HTML]{EFFBEC}\textbf{5} \\
    \bottomrule
    \end{tabular}
}
}
\label{tab:compressions}
\end{table*}

\subsection{Comparasion with State-of-The-Arts (RQ1)}
\label{subsec:sota}
This section uses various text-to-video retrieval models as baselines. 
First, we compare the retrieval performance of HCQ with these baselines. 
Then, we compare HCQ with two state-of-the-art methods, in terms of computation efficiency and storage.

\subsubsection{Performance}
We summarize the ranking performance of retrieval models in Table~\ref{tab:MSRVTT}, Table~\ref{tab:LSMDC} and Table~\ref{tab:ActNet}. 
The results show that HCQ can achieve competitive results with state-of-the-arts. 
Specifically, it outperforms most of the compared methods on MSRVTT and LSMDC. 
Though it does not consistently surpass MMT~\cite{MMT} on ActivityNet Captions, the results are comparable. 
This is partly because we can only leverage motion and audio features on this dataset, which limits the strength of HCQ. 
Besides, we can get two findings from these results. 
(\textbf{i}) Multi-modal cues facilitates better video understanding.
Note that the best performers, including the strongest baselines: MMT~\cite{MMT}, CE~\cite{CE} and HSE~\cite{HSE}, leverage multi-modal information for video representations. 
It demonstrates the promise of leveraging more modalities than vision for video retrieval. 
(\textbf{ii}) Transformer helps to capture multi-modal cues.
Transformer-based methods like the MMT and the proposed HCQ achieve better performance than CE and HSE, showing the advantage of transformers on multi-modal modelling. 

\subsubsection{Efficiency}
Efficiency is an important target for Web search, since it highly relates to the \emph{scalability} of the search engines. 
Here we show results about this rarely investigated aspect in text-video retrieval.
We evaluate text-to-video retrieval with 5 popular dual-encoder retrieval models: \textbf{Dual Encoding}~\cite{DualEncoding}, \textbf{HGR}~\cite{HGR}, \textbf{\citet{mithun2018learning}}, \textbf{CE}~\cite{CE} and \textbf{MMT}~\cite{MMT}, on MSRVTT dataset. 
The manners of these models are as follows: 
    (\textbf{i}) Dual Encoding represents an instance by a 2048-dimensional embedding;
    (\textbf{ii}) HGR represents an instance by 3 1024-dimensional embeddings;
    (\textbf{iii}) Mithun et al. represent an instance by 2 1024-dimensional embeddings;
    (\textbf{iv}) CE represents an instance by 9 768-dimensional embeddings and a 9-dimensional weighting vector;
    (\textbf{v}) MMT represents an instance by 7 512-dimensional embeddings and a 7-dimensional weighting vector;
    (\textbf{vi}) The proposed HCQ represents an instance by 8 256-bit codes. A 256-bit code can be encoded by 32 bytes for compact storage. 
Since MSRVTT only contains 10k video, we duplicate the videos to simulate a large database. 
We evaluate: (\textbf{i}) the average query time, including the time for text encoding time on GPU and nearest neighbor search on CPU; and (\textbf{ii}) the storage overhead for offline-computed video representations. 
We experiment with a NVIDIA GeForce GTX 1080 Ti (11GB) and Intel® Xeon® CPU E5-2650 v4 @ 2.20GHz (48 cores). 
The results are shown in Fig.\ref{fig:efficiency}, demonstrating high query speed and low storage (memory) overhead of HCQ.

\begin{figure}[t]
    \centering
    \includegraphics[width=\linewidth]{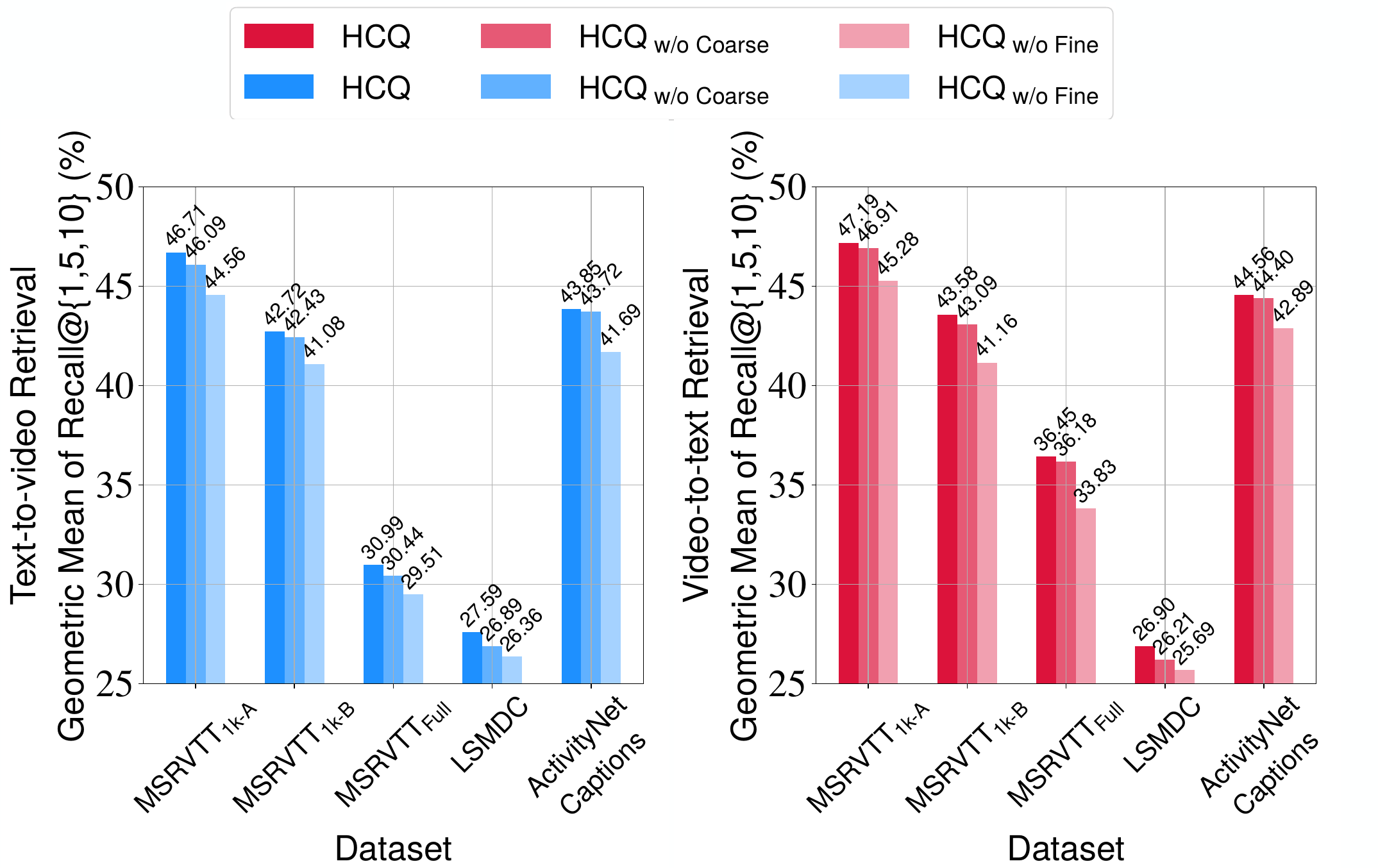}
    \caption{Ablation study on all datasets. Contrastive learnings at coarse-grained and fine-grained levels complement each other, demonstrating improvement on retrieval.}
    \label{fig:ablation}
\end{figure}

\subsection{Comparasion with Other Embedding Compression Approaches (RQ2)}
\label{subsec:compressions}
In this section, we benchmark various hashing methods on text-to-video retrieval. 
First, we construct Hybrid Contrastive Transformer (HCT) by removing the quantization modules of HCQ. 
We adopt embedding-based hybrid-contrastive learning to train HCT and extract dense embeddings.
HCT can serve as a reference model for the upper-bound performance. 
Then we apply post-processing hashing methods to generate compressed representations and conduct retrieval. 
Furthermore, we remove the fine-grained part (\ie the GhostVLAD module) of HCT and construct a Dual-Transformer (DT) that only produces global and coarse-grained embeddings. 
We equip the joint-learning baselines with DT. 
In this way, we adapt these baselines from similar hashing applications, \eg cross-modal image hashing~\cite{DCMH} and text dense retrieval~\cite{JPQ} to our task. 
The reason for using DT rather than HCT is that these models only focus on global representation in their initial works.

The results are shown in Table~\ref{tab:compressions}, which show consistent superiority of HCQ over other embedding compression approaches. 
According to the results, we can get the following findings:
(\textbf{i}) Jointly learning encoder and quantization is effective to mitigate the performance decay. 
Compared with the best post-compression baseline, OPQ~\cite{OPQ}, HCQ achieves average recall (among R@1, R@5 and R@10/R@50) increases of \textbf{1.7}\%, \textbf{2.6}\% and \textbf{1.0}\% on MSRVTT, LSMDC and ActivityNet Captions, respectively. 
(\textbf{ii}) The hybrid-grained contrastive learning provides a powerful supervision the representation models.
We notice that joint-learning baselines with DT perform relatively poorer in our benchmark. 
Compared with DT+DCMH~\cite{DCMH}, HCQ achieves average recall increases of \textbf{6.7}\% and \textbf{6.5}\% on MSRVTT and LSMDC datasets. 
Compared with DT+JPQ~\cite{JPQ}, HCQ achieves average recall increases of \textbf{4.5}\% on ActivityNet Captions dataset.
The reason for large gains is that HCQ learns both coarse-grained and fine-grained quantization by contrastive learning, where the learning at different levels can mutually promote.

\subsection{Ablation Study (RQ3)}
\label{subsec:ablation}
To analyse the effectiveness of hybrid-level contrastive learning, we construct 2 HCQ variants: 
(\textbf{i}) HCQ$_\text{w/o Coarse}$ removes coarse-grained parts of HCQ. It only performs fine-grained alignments and generates fine-grained representations. 
(\textbf{ii}) HCQ$_\text{w/o Fine}$ fine-grained parts of HCQ. It only performs coarse-grained alignments and generates coarse-grained representations. 
The results about the geometric mean of recall@1, recall@5 and recall@10 on all datasets are shown in Fig.\ref{fig:ablation}. According to Fig.\ref{fig:ablation}, removing either coarse-grained part or fine-grained parts of HCQ will result in poorer performance. 
The reason is that, the coarse-grained part provides a easy supervision for corss-view alignment, while the fine-grained part further strengthens it with concept-level information. 
\section{Conclusions}
\label{sec:conclusion}

This paper proposes the first quantized representation learning method, termed Hybrid Contrastive Quantization (HCQ), for efficient cross-view video retrieval. 
We design a hybrid-grained learning strategy to train HCQ, in which we perform Asymmetric-Quantized Contrastive Learning (AQ-CL) at coarse-grained and multiple fine-grained levels to align two views. 
Extensive experiments on benchmark datasets demonstrate the competitive performance of HCQ against state-of-the-art non-compressed methods while simultaneously showing high efficiency in storage and computation. 
More importantly, HCQ sheds light on a promising future direction to better balance performance and efficiency.

\begin{acks}
We want to thank the anonymous reviewers and the meta-reviewer for their valuable comments and suggestions. 
This work is supported in part by the National Natural Science Foundation of China under Grant 62171248, the R\&D Program of Shenzhen under Grant JCYJ20180508152204044, and the PCNL KEY project (PCL2021A07).
\end{acks}

\bibliographystyle{ACM-Reference-Format}
\balance
\bibliography{main}

\clearpage
\appendix

\begin{figure}[t]
    \centering
    \includegraphics[width=\linewidth]{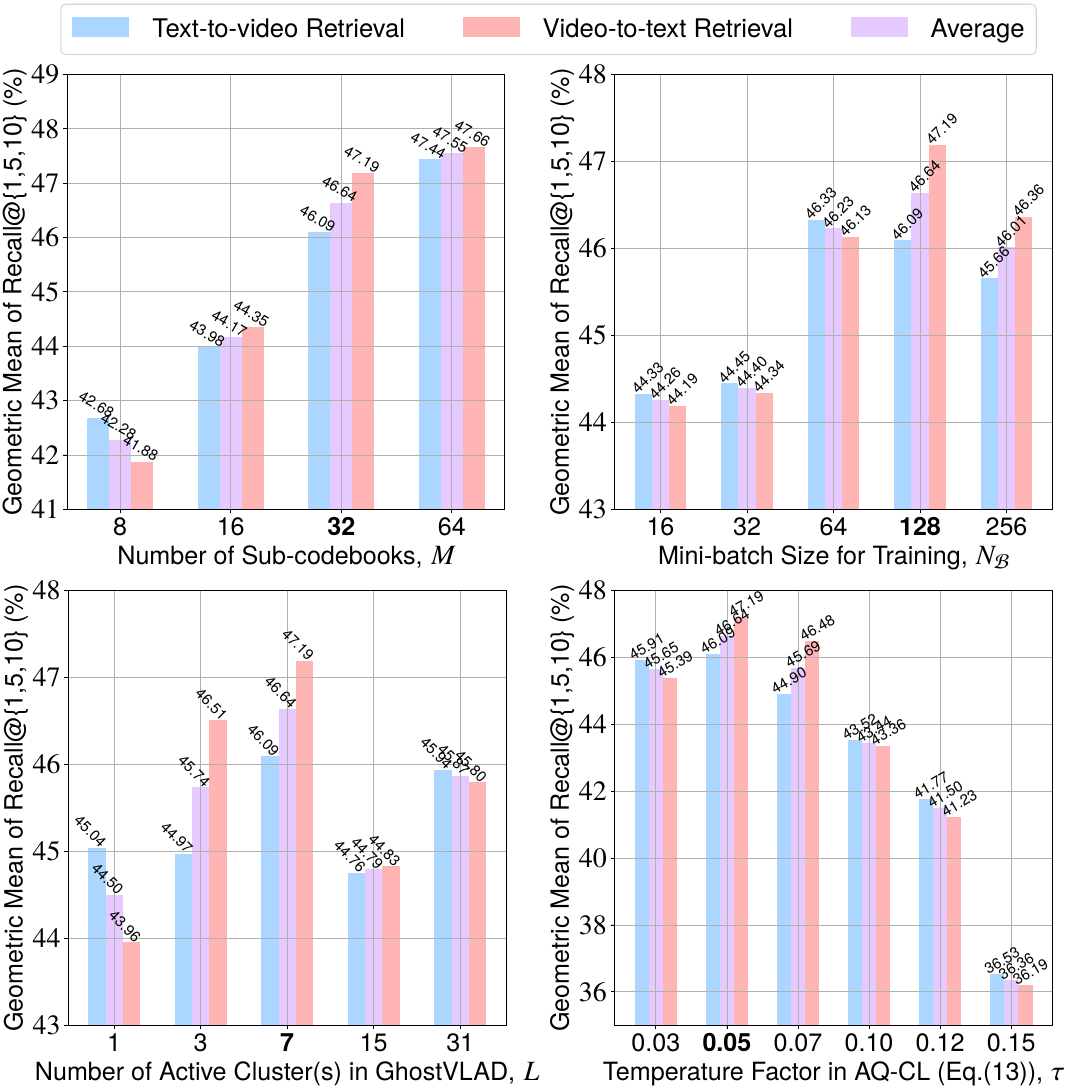}
    \caption{The parameter sensitivity on MSRVTT 1k-A. The default settings (\eg 32 for $M$) are in bold.}
    \label{fig:hyperparameters}
\end{figure}

\section{Appendix: Component Analysis}
\subsection{Parameter Sensitivity}
In this section, We further analyse the sensitivities of different hyper-parameters in HCQ.
The results on the MSRVTT dataset~\cite{MSRVTT} (``1k-A'' split) are shown in Fig.\ref{fig:hyperparameters}.
Based on the results, we can make following brief conclusions:
\begin{enumerate}[label=(\textbf{\roman*}),leftmargin=2em]
    \item $M$ is the sub-space number in each quantization module. Generally, more sub-codebooks yield lower quantization error and better performance, while increasing computational and storage overheads. $M=32$ achieves a reasonable balance;
    \item Both 64 and 128 are good for batchsize on MSRVTT dataset;
    \item $L$ indicates the number of latent concepts. As it rises from $1$ to $7$, the performance first increases because it supports representing more concepts. When $L>7$, the performance drops due to concept redundancy.
    \item $\tau$ controls the penalty intensity of contrastive learning, which is sensitive. $\tau=0.05$ on MSRVTT dataset can produce satisfactory performance.
\end{enumerate}

\begin{table}[t]
\caption{We investigate the effect of adopting different BERT variants on MSRVTT dataset~\cite{MSRVTT} (``1k-A'' split). The default initialization for text encoder is ``BERT-base-cased''.}
\centering
\resizebox{\columnwidth}{!}{
\setlength{\tabcolsep}{0.2em}{
\begin{tabular}{lccccccccc}
\toprule
 & \multicolumn{4}{c}{Text $\rightarrow$ Video} & & \multicolumn{4}{c}{Video $\rightarrow$ Text} \\
\cmidrule{2-5} \cmidrule{7-10}
\multirow{-2}{*}{Method} & R@1$\uparrow$ & R@5$\uparrow$ & R@10$\uparrow$ & MdR$\downarrow$ & & R@1$\uparrow$ & R@5$\uparrow$ & R@10$\uparrow$ & MdR$\downarrow$ \\
\toprule
\rowcolor[HTML]{EFFCEB} HCQ & 25.9 & 54.8 & 69.0 & 5 && 26.3 & 57.0 & 70.1 & 4 \\
HCQ$_\text{ BERT-large}$ & 27.4 & \textbf{57.7} & \textbf{70.7} & \textbf{4} && 26.2 & \textbf{59.0} & \textbf{71.8} & 4 \\
HCQ$_\text{ DistilBERT-base}$ & 25.4 & 54.2 & 67.3 & \textbf{4} && 26.3 & 56.4 & 69.0 & 4 \\
HCQ$_\text{ RoBERTa-base}$ & 25.5 & 54.7 & 67.8 & 5 && 24.5 & 55.0 & 69.0 & 4 \\
HCQ$_\text{ RoBERTa-large}$ & \textbf{28.0} & 55.4 & 68.5 & \textbf{4} && \textbf{27.0} & \textbf{59.0} & 68.4 & 4 \\
HCQ$_\text{ XLNet-base}$ & 25.8 & 56.2 & 68.7 & 5 && 24.6 & 55.5 & 69.0 & 4 \\
HCQ$_\text{ XLNet-large}$ & 25.0 & 53.0 & 66.6 & 5 && 25.3 & 54.5 & 68.0 & 4 \\
\bottomrule
\end{tabular}}}
\label{tab:bert_ablation}
\end{table}

\subsection{Adopting Other BERT Variants}
During our submission to the WWW'22, a valuable question from the anonymous reviewer \#2 raises our interest in leveraging other BERT variants to initialize the text encoder. 
We have done experiments with ``BERT-large''~\cite{BERT}, ``DistillBERT-base''~\cite{DistilBERT}, ``RoBERTa-base''~\cite{RoBERTa}, ``RoBERTa-large''~\cite{RoBERTa},  ``XLNet-base''~\cite{XLNet} and ``XLNet-large''~\cite{XLNet} based on the Hugging Face (\url{https://huggingface.co/models}) implementation.
The results of these trials are shown Table~\ref{tab:bert_ablation}. Large version of BERT models, \eg ``BERT-large'' and ``RoBERTa-large'', help to improve retrieval performance. 

\end{document}